\documentstyle[12pt,aasms4]{article}
\begin{document}
\title{A QSO in a cluster at z=2.4}
 
\author{J.B.Hutchings\altaffilmark{1}}
\affil{Dominion Astrophysical Observatory, National research Council of
Canada, 5071 W. Saanich Rd., Victoria B.C. V8X 4M6, Canada}
\author{S.G.Neff\altaffilmark{1}}
\affil{Laboratory for Astrophysics, Code 600, NASA Goddard Space Flight
Center, Greenbelt, MD20771}
\authoremail{hutchings@dao.nrc.ca,neff@stars.gsfc.nasa.gov}
\altaffiltext{1}{Guest Observer, Canada France Hawaii Telescope,
which is operated by NRC of Canada, CNRS of France, and the University 
of Hawaii}

\begin{abstract}

Photometry is presented in R, J, H, K of the z=2.37 QSO 0820+296 and its environment.
Colours and magnitudes of most of the objects within 50 arcsec of the QSO suggest 
they are a group of galaxies at the QSO redshift with a young stellar population. 
\end{abstract}

\section{Introduction and data}

The QSO 0820+296 is reported as having redshift 2.369 by Wills and Wills (1976),
and Carswell et al (1976), and to have absorptions shortward of the emission lines 
of Ly$\alpha$ and C IV. It is radio-loud with 5 Ghz flux of 400 mJy 
(Pauliny-Toth et al 1972). It was listed in the Hewitt and Burbidge 1987 catalogue 
but omitted from their 1993 catalogue, without any explanation. 
It is listed in the Veron and Veron catalogue (1991). We find no reference in 
other literature to suggest that the QSO redshift is in doubt. The QSO has magnitude 
v=18.5 which corresponds to M$_V$ = -27.1 without k-correction. 

R-band imaging was obtained at the CFHT HR Cam, as reported by Hutchings (1995a,b)
with $\sim$0.1 arcsec pixels and 0.8 arcsec FWHM images. These data indicated 
a 3$\sigma$ excess of faint galaxy counts in a 1 arcmin box around the QSO, and 
also that the QSO itself is 
marginally resolved. We have since obtained deep J, H, and K band
images with the Redeye Wide camera of the CFHT, in order to use
colour photometry to constrain the properties of the faint galaxies.

The J,H,K exposures were obtained in January 1996, using a 120 arcsec field with 
0.5 arcsec pixels.
The observations were taken using a dither pattern with 10" offsets,
giving a uniformly exposed field of 90". Total integration was for 35 minutes
in each band. Image quality was about 0.9" so that the images are
not well sampled. However, the large pixels enabled good S/N within
the faint field galaxies. 

The data were processed as described in more detail by Hutchings
and Neff (1997: HN), and all faint objects in the field measured.
We retained for discussion only those with measuring errors 0.15
mag or better in all 3 NIR bands. As discussed in HN, this occurs at
J$\sim$21, H$\sim$20, K$\sim$19.5. This yielded 24 objects,
of which 3 are clearly seen to be stellar in the well-sampled R-band image.
These stars also have NIR colours characteristic of z=0 stellar
populations and are not included in the subsequent discussion.
Figure ~\ref{image} shows the J-band image with the measured objects marked.
There are some fainter objects not marked, particularly within 30" of the QSO.

The photometry was done using the imexam task in IRAF\footnote{IRAF is distributed 
by NOAO, which is operated by AURA Inc., under cooperative agreement with 
the NSF.} using several settings
of radius and background annulus, removing any hot pixels or
nearby objects individually. The radii used were large enough to contain
azimuthally averaged signal down to the noise level. Because of slightly different
field placement, galaxies 19, 20, and 21 were not in the R-band image.
Table 1 shows the photometric measures used.

\section{Results}

   Figure~\ref{ccd} shows some 2-colour plots of the marked objects in Figure 1.
We discuss below the close clumping of the colours of most of these galaxies.
The H-K index is not preferred (or shown) as it involves the poorest precision
measures (K) with the shortest colour spread, and the K band needs 
correction from the models which use K while our observing bandpass
was K', calibrated with K standard magnitudes. Nevertheless, the
objects are still remarkably clumped in that index too.

   Figure~\ref{ccd} also includes Gissel model tracks (Bruzual and Charlot 1993)
for stellar populations. The tracks are for populations with an initial starburst 
of duration 1 Gyr, followed by passive evolution. Tracks for continued star-formation
or exponentially declining
star-formation follow essentially the same tracks but remain nearer the
initial point of the models shown. The diagram marks the tracks at ages 10$^8$yr
and then in units of Gyr after the initial starburst.

  The plots may be compared with similar results on low z QSOs by HN,
which also show a population of associated galaxies with the QSOs, 
but which also have a number of non-associated (mainly background) galaxies.
 In low redshift fields HN noted that there may be
systematic errors in the models, as they are uncertain at NIR rest
wevelengths.
However, the NIR models at high redshifts reflect visible range rest
wavelengths, and above z=0.4 (Hutchings and Davidge 1996), 
the models appear to match the observations
very well. Thus, we consider the model comparisons in this paper to be good.  

   Figure~\ref{ccd} is remarkable in the tight clumping of the objects. 
The distribution in other high latitude fields from similar data is seen
in HN, and Hutchings and Davidge (1997). Typical faint
galaxies are evenly distributed in R-J and J-K over values from 1 to 3 in each.
The QSO itself is an outlier, corresponding to a very young population, or more
likely, nuclear power-law
continuum, at its redshift. The region populated by the data corresponds
to either a reddened old population at very low redshift, or a 1-2 Gyr old
population at z=2.4. The size and brightness of the galaxies suggests that
they are at large redshift, so that the plot is consistent with, and suggestive
of a group of galaxies with young stellar populations, associated with the QSO.

  The galaxies 5, 17, and 18 lie furthest from the main group in both panels
of Figure~\ref{ccd}. They have the highest R-J values and also lie outside the
region covered by the models. We (somewhat arbitrarily) regard these as outliers
or possibly spurious in some way.  

   As projected on the sky, the QSO lies centrally in the group indicated by
Figure~\ref{ccd} (i.e. without the outliers), and as reported by Hutchings (1995a),
there is an excess of faint galaxies to 1 arcmin around the QSO. Figure~\ref{hist}
shows the distribution of magnitudes of the galaxies and the QSO in 4 colours. 
These are incomplete at faint magnitudes, particularly in J and H, by the 
detection selection described in the previous section. The distance
modulus for the QSO is $\sim$46 (for H$_0\sim$70 cosmologies), 
so the uncorrected absolute magnitudes
are -23 and brighter in R, to $\sim$-28 in K. If they are young populations,
this corresponds to M$_R$=-21. The R-J colours are plotted against
magnitude in Figure~\ref{cmd}, with the R-J colour evolution of young populations
at redshifts 2 and 3. The outliers in Figure~\ref{ccd} are those with the highest
R-J colour. If these galaxies are not in the QSO group, there is no colour-magnitude
correlation, so that the galaxies appear to have the same ages regardless
of their brightness. The R-J colours suggest that the galaxies are in an
early post-starburst stage, but we cannot tell if they are initial starbursts
or a strong second starburst on top of an older population. They do not correspond
to the oldest population that might exist at z=2.4 ($\ge$2.5Gyr).

  The ten `best' companions (i.e. tightest grouping in Figure 2)
are found within an
area projected as 50 arcsec in the sky, which is 250Kpc at z=2.4. This
is a very high density of bright galaxies if they are all associated.
Hutchings (1995) reported increased galaxy counts near all of a small sample
of z=2.4 QSOs. The excess over the local background in the case of 0820+296 
is a factor of two: to a similar limiting R magnitude this is $\sim$10 galaxies 
in a diameter of 50" (Hutchings 1995a). 
Using similar colour photometry, we have found a similar
collection of galaxies around the z=2 QSO first discussed by Dressler et al
(1994): see Hutchings and Davidge (1996). Pascarelle et al (1996) have recently
confirmed a high density of faint companion galaxies
in the field of a z=2.4 radio galaxy. Thus, there is mounting evidence that high z 
AGN are found in compact groups of very blue galaxies that are apparently
in an early starburst stage. 

   The spectrum of 0820+296 is reported to have absorption systems that
may be near z=2. Galaxy 5 has colours of a somewhat more evolved population
at redshift 2 and lies very close to the QSO, so might
be responsible. Clearly, spectroscopy is needed to investigate these results
more definitively. At z=2.4, the strong stellar wind lines of OB stars should 
be in the visible range.

\clearpage
{\bf References}

Bruzual G., and Charlot S., 1993, ApJ, 405, 538

Carswell R.F. et al 1976, A+A, 53, 275

Dressler A., Oemler A., Gunn J.E., Butcher H., 1993, ApJ, 404, L45

Hewitt A. and Burbidge G.R., 1987, ApJS, 63, 1

Hewitt A. and Burbidge G.R., 1993, ApJS, 87, 451

Hutchings J. B. 1995a, AJ, 109, 928

Hutchings J.B. 1995b, AJ, 110, 994

Hutchings J. B. and Davidge T. 1997, PASP submitted

Hutchings J. B., and Neff S. G., 1997, AJ, in press: HN

Pascarelle S.M., Windhorst R.A., Keel W.C., Odewahn S.C., 1996, Nature, 383, 45

Veron-Cetty M-P and Veron P. 1991, ESO Scientific Report 10

Wills B.J., 1976, AJ, 81, 1031

Wills D., and Wills B.J., 1976, ApJS, 31, 143

\clearpage
\centerline{Captions to figures}

\figcaption[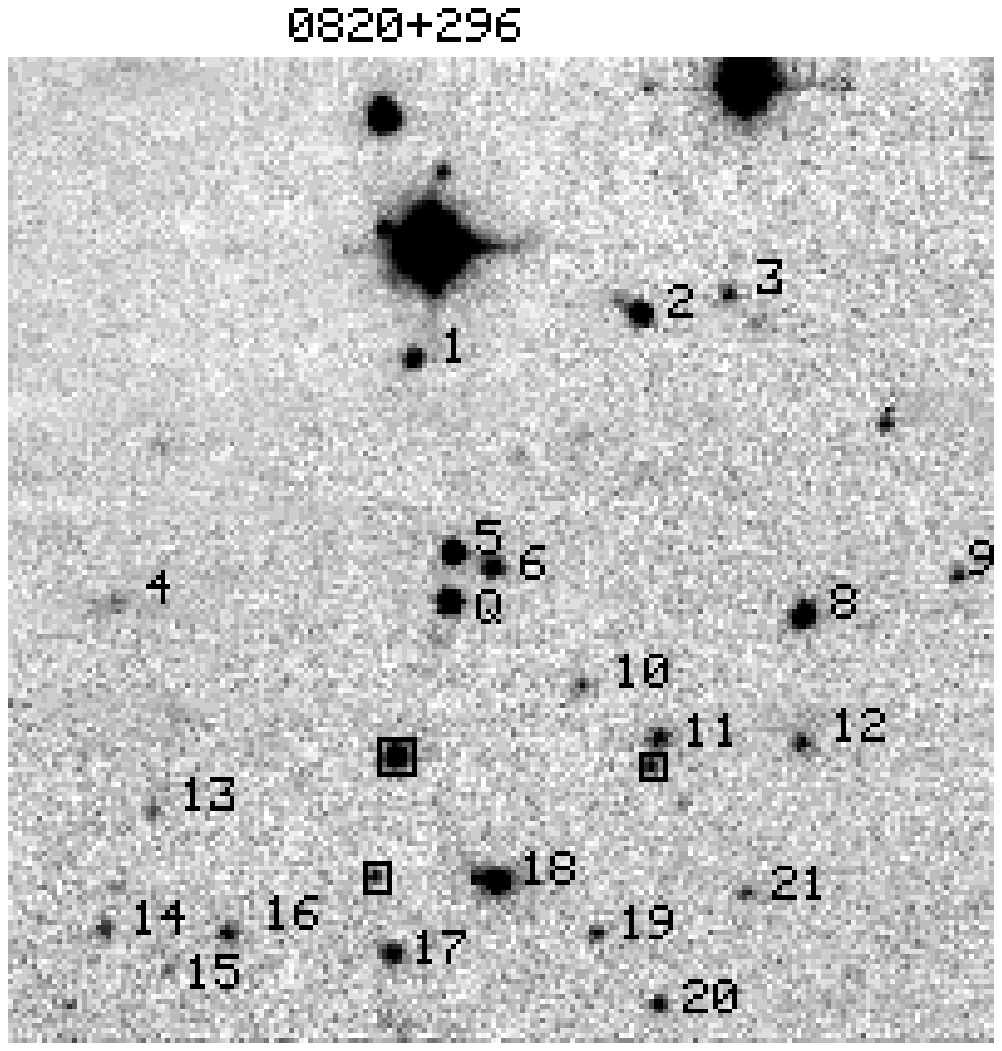]{J-band image of 90 arcsec field near 0820+206.
North is left and East is down. Stars are enclosed in squares. 
The QSO is marked Q. Galaxies with photometric errors $<$0.15mag in
J, H, and K are numbered. \label{image}}

\figcaption[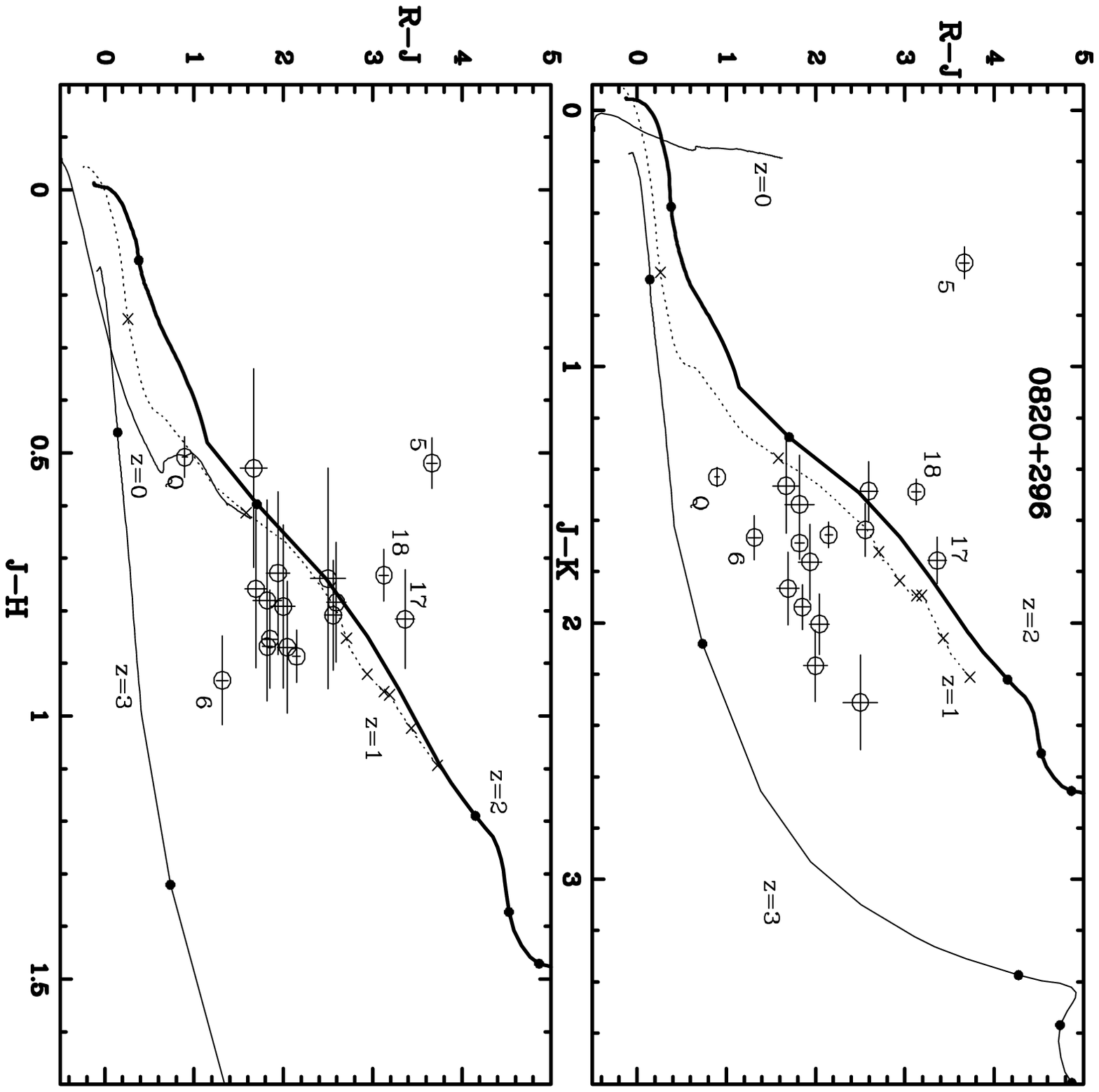]{Two-colour plots of measured galaxies in Figure 1.
Lines are evolution tracks (all starting at lower left) for populations at redshifts
as marked. The populations all have an initial 1 Gyr starburst followed by passive
evolution. Dots or crosses along the tracks mark 10$^8$yr and then 1 Gyr intervals 
- galaxies move
faster across the diagram as redshift increases. Extended or renewed star-formation
keeps the values closer to the starting points. Note the close clumping of
most objects in the diagrams, at the values expected for post-starburst galaxies
at redshift near 2.4. 5, 17, 18 are (possibly foreground) outliers, and 6 lies
very close to the QSO. The QSO itself has smaller colour indices, probably due to
its nuclear SED. \label{ccd}}

\figcaption[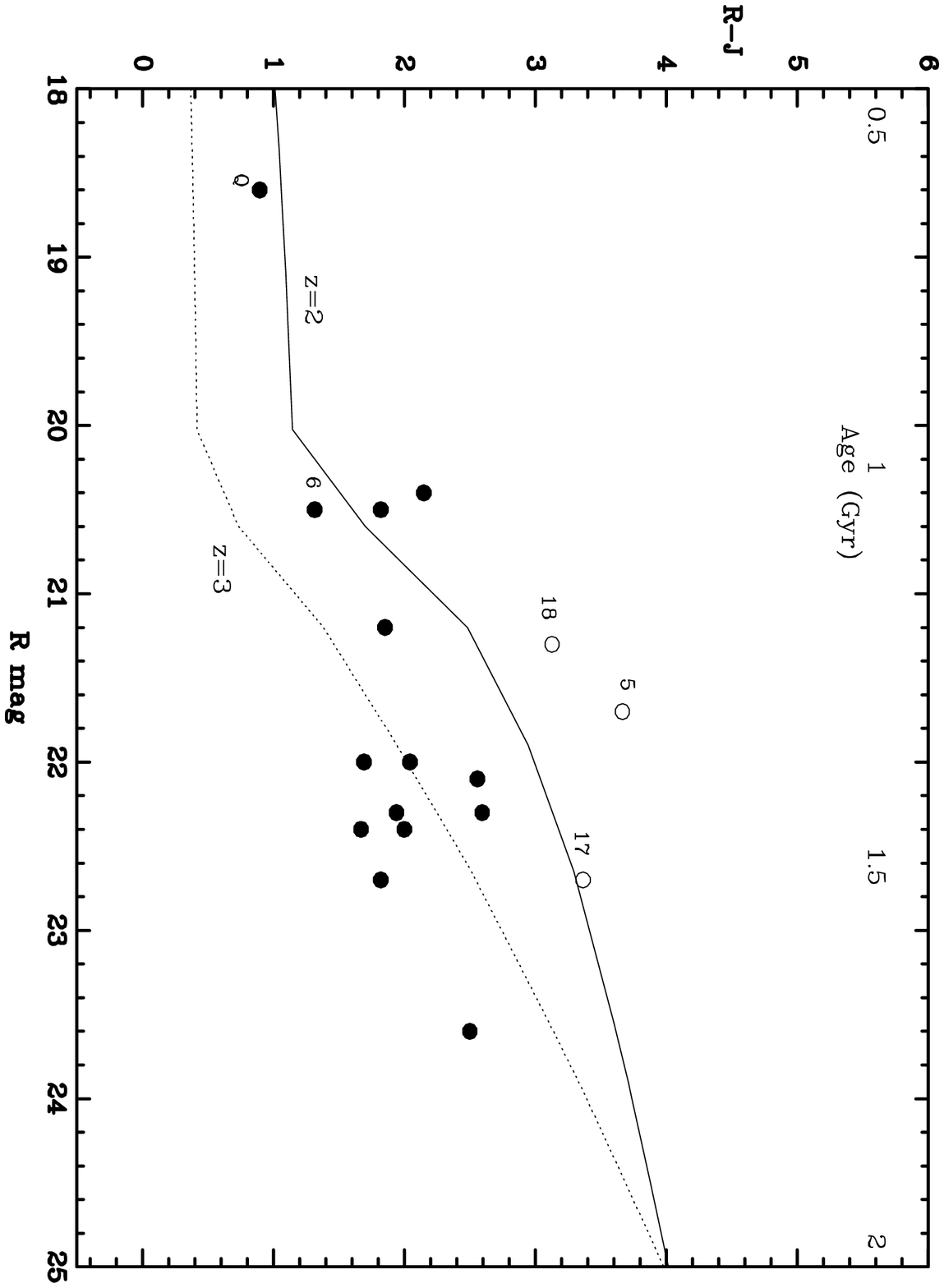]{Colour-magnitude plot of galaxies in Figure 1. The
outliers 5, 17, 18 have open symbols. The remainder have a small range of R-J
that is not correlated with magnitude. The lines show model colours at the redshifts
shown, with ages indicated at the top. The galaxy positions all correspond to 
populations that are just beyond an initial starburst, 
or are undergoing extended or renewed star-formation. \label{cmd}}

\figcaption[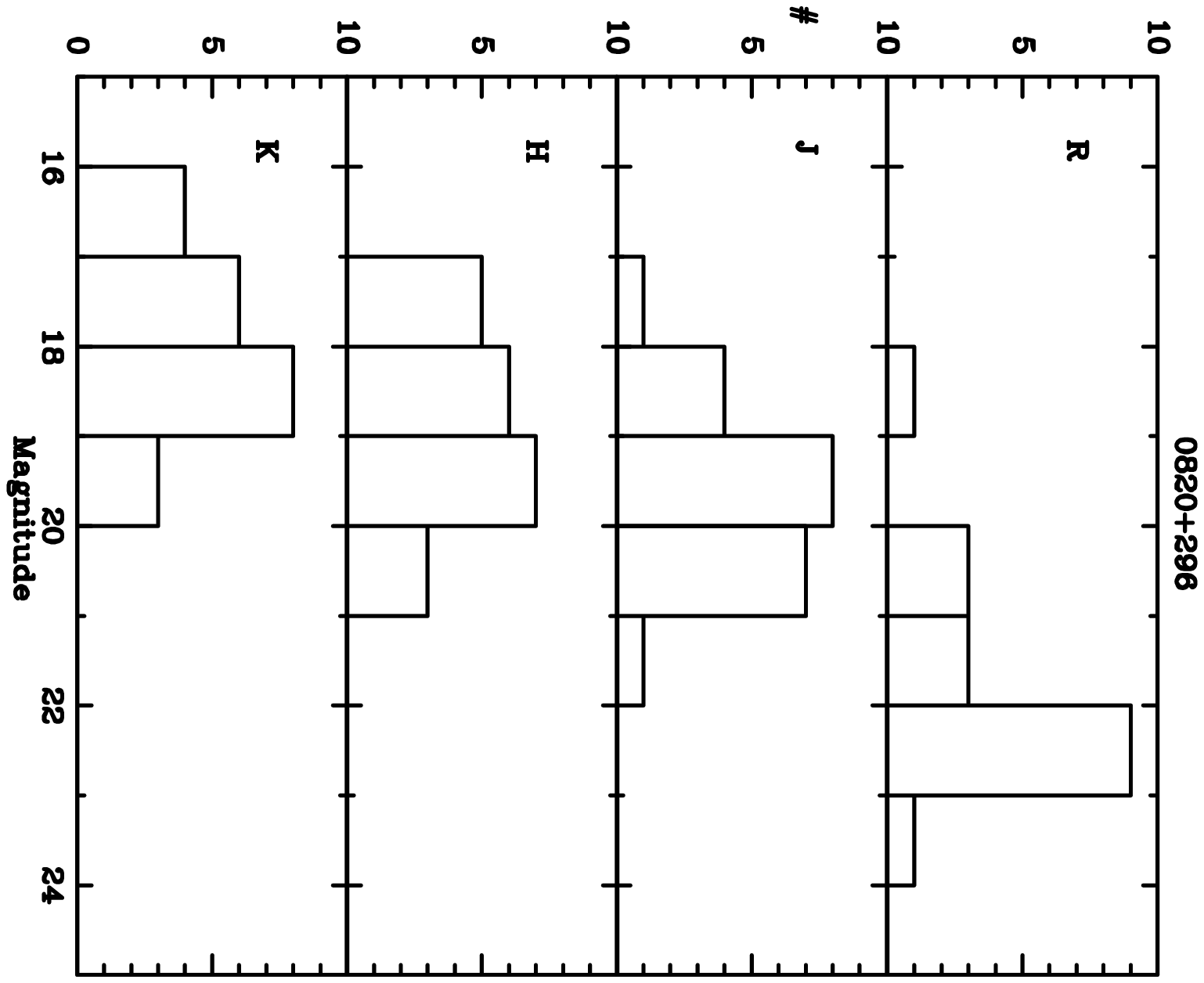]{Histograms of magnitudes in 4 colours in the
0820+296 field, used in the discussion. The QSO is in the brightest bin, and the
faint ends are incomplete, particularly in H and K, by our measuring-error
selection. \label{hist}}
\end{document}